\documentclass[conference]{IEEEtran}
\usepackage{amssymb}
\usepackage[usenames,dvipsnames,svgnames,table]{xcolor}		
\usepackage{url}		
\usepackage{epstopdf}

\usepackage{fixme}
\fxsetup{
  status=draft, 
  multiuser,
  theme=color,
  inlineface=\bfseries,
  envface=\bfseries
}
\FXRegisterAuthor{ne}{ane}{NE} 

\usepackage{cite}

\ifCLASSINFOpdf
   \usepackage[pdftex]{graphicx}
   \graphicspath{{../eps/}}
   \DeclareGraphicsExtensions{.eps,.pdf,.jpeg,.png}
\else
   \usepackage[dvips]{graphicx}
   \graphicspath{{../eps/}}
\fi
\usepackage{caption}
\usepackage[font=footnotesize]{subfig}
\usepackage{url}

\begin{document}
%
\title{The \mbox{DEEP-ER} project: \\ I/O and resiliency extensions for the Cluster-Booster architecture}


\author{\IEEEauthorblockN{Anke Kreuzer, Norbert Eicker, Estela Suarez*}
\IEEEauthorblockA{J\"{u}lich Supercomputing Centre (JSC) \\ 
		Institute for Advanced Simulation (IAS)\\
        Forschungszentrum J\"{u}lich GmbH \\  
		52425 J\"{u}lich, Germany \\
	Email*: e.suarez@fz-juelich.de}	
\and
\IEEEauthorblockN{Jorge Amaya}
\IEEEauthorblockA{Katholieke Universiteit Leuven \\ 
	BE-3001, Heverlee, Belgium}
\and
\IEEEauthorblockN{Raph\"{a}el L\'{e}ger}
\IEEEauthorblockA{INRIA Sophia-Antipolis M\'{e}diterran\'{e}e \\ 
	Sophia-Antipolis, France}
}


%


\maketitle

\begin{abstract}
The recently completed research project \textit{\mbox{DEEP-ER}} has 
developed a variety of hardware and software technologies
to improve the I/O capabilities of next generation
high-performance computers, and to enable applications
recovering from the larger hardware failure rates expected 
on these machines.

The heterogeneous Cluster-Booster architecture 
--~first introduced in the predecessor DEEP project~-- 
has been extended by a multi-level
memory hierarchy employing non-volatile and network-attached 
memory devices. Based on this hardware infrastructure, 
an I/O and resiliency software stack has been implemented 
combining and extending well established libraries and software tools, 
and sticking to standard user-interfaces. 
Real-world scientific codes have tested the 
projects' developments and demonstrated the improvements 
achieved without compromising the 
portability of the applications. 
\end{abstract}


%
\IEEEpeerreviewmaketitle

\section{Introduction}
\label{sec:intro}
During the last decade
the computing performance of HPC systems is growing much faster 
than their memory bandwidth and capacity~\cite{McKee:2004}. 
This so-called \textit{memory wall} is not expected to disappear 
in the near future.
Additionally, higher failure rates are predicted for next generation 
machines due to their huge number of components. 
These are two of the main issues most directly affecting
the scientific throughput that can be extracted from Exascale systems.

The \textit{DEEP projects}~\cite{DEEPweb} are a series 
of (by now) three EC funded projects (DEEP, \mbox{DEEP-ER}, 
and \mbox{DEEP-EST}) that address the 
Exascale computing challenges with their research. 
All three follow a stringent \textit{co-design} 
strategy, in which full-fledged scientific applications 
guide the design and implementation of system 
hardware and software. Their requirements, identified 
by detailed application analysis, guide  
all the projects' developments. The selected codes 
have also been adapted to the project platforms and 
served as a yardstick to validate and benchmark the hardware and
software achievements implemented in the course of the projects.

This paper describes the technology developed within 
the \mbox{DEEP-ER} project to improve the I/O and 
resiliency capabilities of HPC systems. 
In particular, the heterogeneous Cluster-Booster 
architecture~\cite{eicker:PARS} --~introduced in DEEP~--
was extended by a multi-level memory hierarchy.
\ This served as a
foundation of a complete I/O and resiliency software stack. 

Section~\ref{sec:arch} of this paper
presents the \mbox{DEEP-ER} system architecture, 
including the underlying Cluster-Booster concept, 
the specific hardware configuration of the 
\mbox{DEEP-ER} prototype, 
and its memory hierarchy and technologies. 
The software stack is explained in Section~\ref{sec:sw}, 
including the programming environment 
already introduced in the predecessor DEEP project, 
and --~more detailed~-- the \mbox{DEEP-ER} 
I/O and resiliency software developments.
The co-design applications are shortly described 
in Section~\ref{sec:apps}. A selection of results 
obtained during the evaluation of the \mbox{DEEP-ER} 
concepts are presented in Section~\ref{sec:results},
while Section~\ref{sec:relwork} puts them in context 
with related work. Finally, the conclusions of the 
paper are summarized in Section~\ref{sec:concl}.

\section{System Architecture}
\label{sec:arch}
Cluster computing enables building high-performance systems 
benefiting from lower cost of commodity of the shelf (COTS) components. 
Traditional, homogeneous clusters are built by connecting a 
number of general purpose processors (e.g. Intel Xeon, 
AMD Opteron, etc.) by a high speed network (e.g. InfiniBand or
OmniPath). This approach is limited by the relatively 
high power consumption and cost per performance of general 
purpose processors. Both make a large scale
homogeneous systems extremely power hungry and costly. 

The cluster's overall energy and cost efficiency can be improved by 
adding accelerator devices (e.g. many-core processors or 
general purpose graphic cards, GPGPUs), which provide 
higher Flop/s performance per Watt. Standard heterogeneous 
clusters are built attaching one or more accelerators 
to each node. However, this \textit{accelerated 
node} approach presents some caveats. An important one is 
the combined effect of the accelerators' dependency on the 
host CPU and the static arrangement of hardware resources, 
which limits the accessibility of the accelerators for other 
applications than the one running on the host CPU. Furthermore, both
CPU and accelerator have to compete for the limited network
bandwidth in this concept.

\subsection{Cluster-Booster concept}
\label{sec:CBconcept}
\begin{figure}[!t]
\centerline{
\subfloat[Sketch of the Cluster-Booster
  architecture as implemented in the \mbox{DEEP-ER} project 
  (KNL: Knights Landing; NVM: non-volatile memory;
  NAM: network attached memory).]
  {\includegraphics[height=4cm]{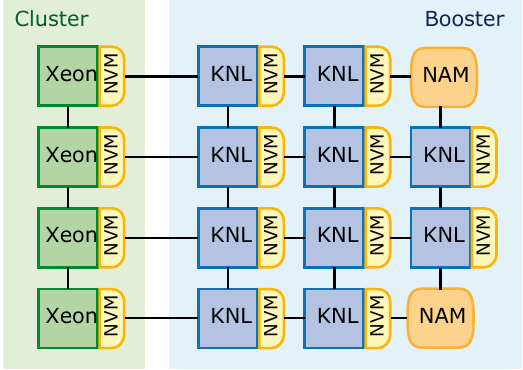}
\label{fig:CBarch_scheme}}
\hfil
\subfloat[Picture of the \mbox{DEEP-ER} prototype, at JSC.]
  {\includegraphics[height=4cm]{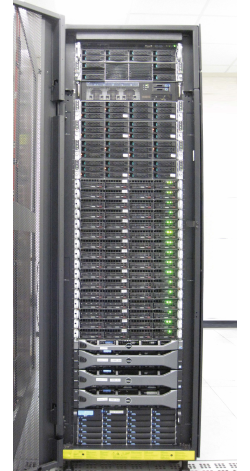}
\label{fig:deeper_proto}}}
\caption{Cluster-Booster architecture in \mbox{DEEP-ER}.}
\label{fig:CBarch}
\end{figure}

The \textit{Cluster-Booster architecture} 
(\figurename~\ref{fig:CBarch_scheme}) integrates heterogeneous computing 
resources at the system level. Instead of plugging accelerators 
into the node and attaching them directly to the CPUs, 
they are moved into a stand-alone
cluster of accelerators that has been named \textit{Booster}.
It is capable to run full codes 
with intensive internal communication, leveraging the fact that
accelerators therein are 
autonomous and do communicate directly with each 
other through a high-speed network without the help of an additional CPU. 

%

The Booster is attached to a standard HPC Cluster 
via a high-speed network. This connection, together 
with a uniform software stack running over both parts 
of the machine (see Section~\ref{sec:sw}), enables 
Cluster and Booster acting together as a unified system. 
This opens up new prospects to application developers,
who have now full freedom to decide how they distribute their 
codes over the system. 
In contrast to accelerated clusters the Cluster-Booster 
concept poses no constraints on the
combined amount of CPU and accelerator nodes that an 
application may select, since resources are reserved 
and allocated independently. Performance benefits of an 
application distributed over Cluster and Booster on the 
\mbox{DEEP-ER} prototype are discussed in~\cite{archPaper}.

\subsection{Prototype hardware configuration}
\label{sec:sysconf}
The first prototype of the Cluster-Booster concept 
was designed and built in the course of the \textit{DEEP} 
project~\cite{DEEPweb}. The later \textit{\mbox{DEEP-ER} prototype} 
(\figurename~\ref{fig:deeper_proto}) is the second 
generation of the same architecture and was installed 
at the J\"{u}lich Supercomputing Centre (JSC) in 2016. 
It consists of 16 Cluster nodes and 8 Booster 
nodes. The system's configuration is detailed in 
Table~\ref{tab:hwconfig}. 
Cluster and Booster modules are integrated into a single,
air-cooled 19'' rack. This rack also holds the storage system (one
meta-data, two storage servers, and 57~TB of storage on spinning disks).
A uniform high-speed Tourmalet~A3 EXTOLL fabric  
runs across Cluster and Booster, connecting 
them internally, among each other, and to the central 
storage~\cite{extoll}.

\begin{table}[t]\scriptsize
  \centering
   \caption{\label{tab:hwconfig} Hardware configuration of the \mbox{DEEP-ER} prototype.}
   \begin{tabular}{l l l}
      \hline
      {\bf Feature} & {\bf Cluster} & {\bf Booster}  \\
      \hline \hline
      Processor & Intel Xeon~E5-2680~v3 & Intel Xeon Phi 7210 \\
      Microarchitecture & Haswell & Knights Landing (KNL) \\
      Sockets per node	& 2 & 1 \\
      Cores per node & 24 & 64 \\
      Threads per node & 48 & 256\\
      Frequency  & 2.5~GHz & 1.3~GHz \\
      \hline
      Memory (RAM) & 128~GB & 16~GB -- MCDRAM \\
      & & 96~GB -- DDR4 \\
      NVMe capacity	& 400~GB & 400~GB \\
      \hline
      Interconnect & EXTOLL Tourmalet A3 & EXTOLL Tourmalet A3 \\
      Max. link bandwidth & 100~Gbit/s & 100~Gbit/s \\
      MPI latency & 1.0~$\mu$s & 1.8~$\mu$s \\
      \hline
      Node count & 16 & 8 \\
      Peak performance & 16~TFlop/s & 20~TFlop/s \\
      \hline
    \end{tabular}%
\end{table}%

\subsubsection{Non-volatile Memory}
\label{sec:nvm}
The \mbox{DEEP-ER} prototype is enhanced by advanced 
memory technologies. A multi-level memory 
hierarchy has been built providing a total memory 
capacity of 8~TByte, to enable the implementation of 
innovative I/O and resiliency techniques 
(see Sections~\ref{sec:io} and \ref{sec:resil}). 
Each node in the \mbox{DEEP-ER} prototype (in both Cluster 
and Booster) feature a non-volatile memory (NVM) device
for efficient buffering of I/O operations and writing checkpoints. 
The chosen technology is Intel's DC~P3700, a device aimed to replace
SSDs, with 400~GByte capacity. It provides high speed, 
non-volatile local memory, attached to the node via 4~lanes 
of PCIe~gen3. 
Extensive experiments and a wide range of
measurements with I/O benchmarks and application mock-ups 
have been performed, which shows substantial performance 
increase over conventional best-of-breed SSDs and 
state-of-the art I/O servers (see Section~\ref{sec:io_results}), 
in particular for scenarios with many I/O requests in parallel.

\subsubsection{Network Attached Memory (NAM)}
\label{sec:nam}

\begin{figure}[!ht]
\centering
\includegraphics[width=0.6\columnwidth]{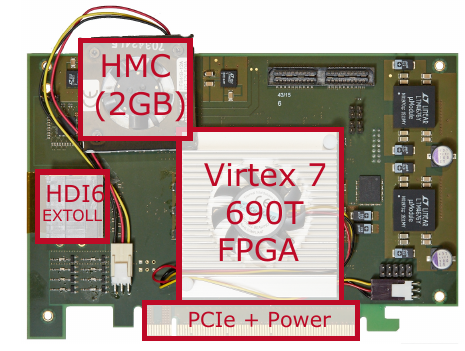}
\caption{Network Attached Memory (NAM) board.}
\label{fig:nam}
\end{figure}

\mbox{DEEP-ER} has also introduced a new
memory concept: the \textit{Network Attached Memory} (NAM)~\cite{schmidtPhD}. 
It exploits the RDMA capabilities of the EXTOLL fabric,
which enable accessing remote memory resources without 
the intervention of an active component (as a CPU) on the remote side.
The NAM board (\figurename~\ref{fig:nam}) is built using a PCIe form-factor 
 and acts fully autonomously. 
The PCIe connection is only utilized for power supply and debugging
purposes.
The NAM combines Hybrid Memory Cube (HMC) resources
with a state-of-the-art Xilinx Virtex~7 FPGA. 
The \mbox{DEEP-ER} prototype holds two NAM devices with a capacity 
of 2~GByte each. This relatively small size is due to current limitations
of the HMC technology. 
The FPGA implements three functions: an HMC controller\footnote{The 
HMC controller has been released as Open Source~\cite{openhmc}.}, 
the EXTOLL network interface with two full-speed Tourmalet 
links, and the NAM logic. Together, they create a
high-speed memory device, which is directly attached 
to the EXTOLL fabric and therefore globally accessible by all 
nodes in the system.



\begin{figure}[!ht]
\centering
\includegraphics[width=0.9\columnwidth]{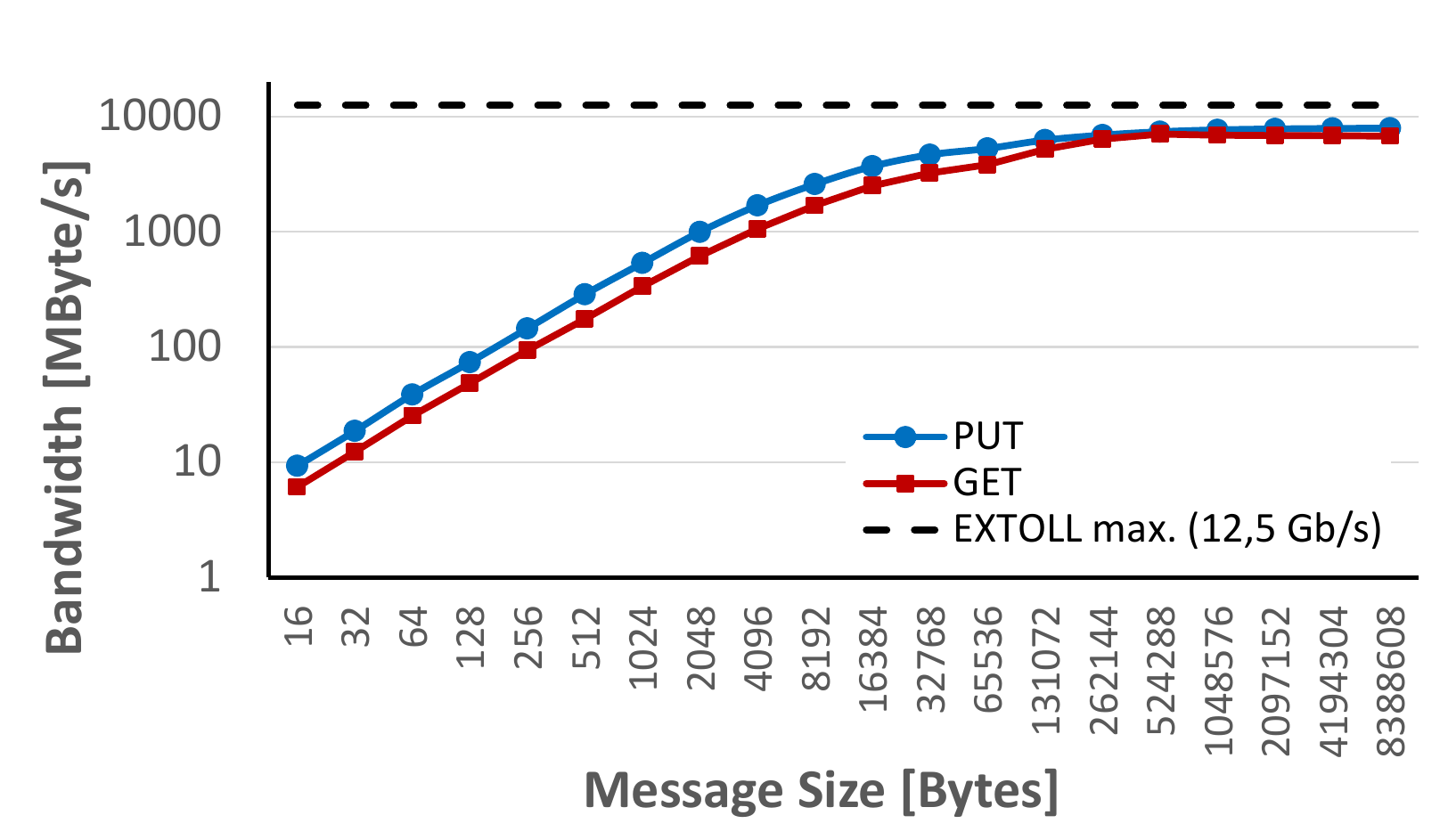}
\includegraphics[width=0.9\columnwidth]{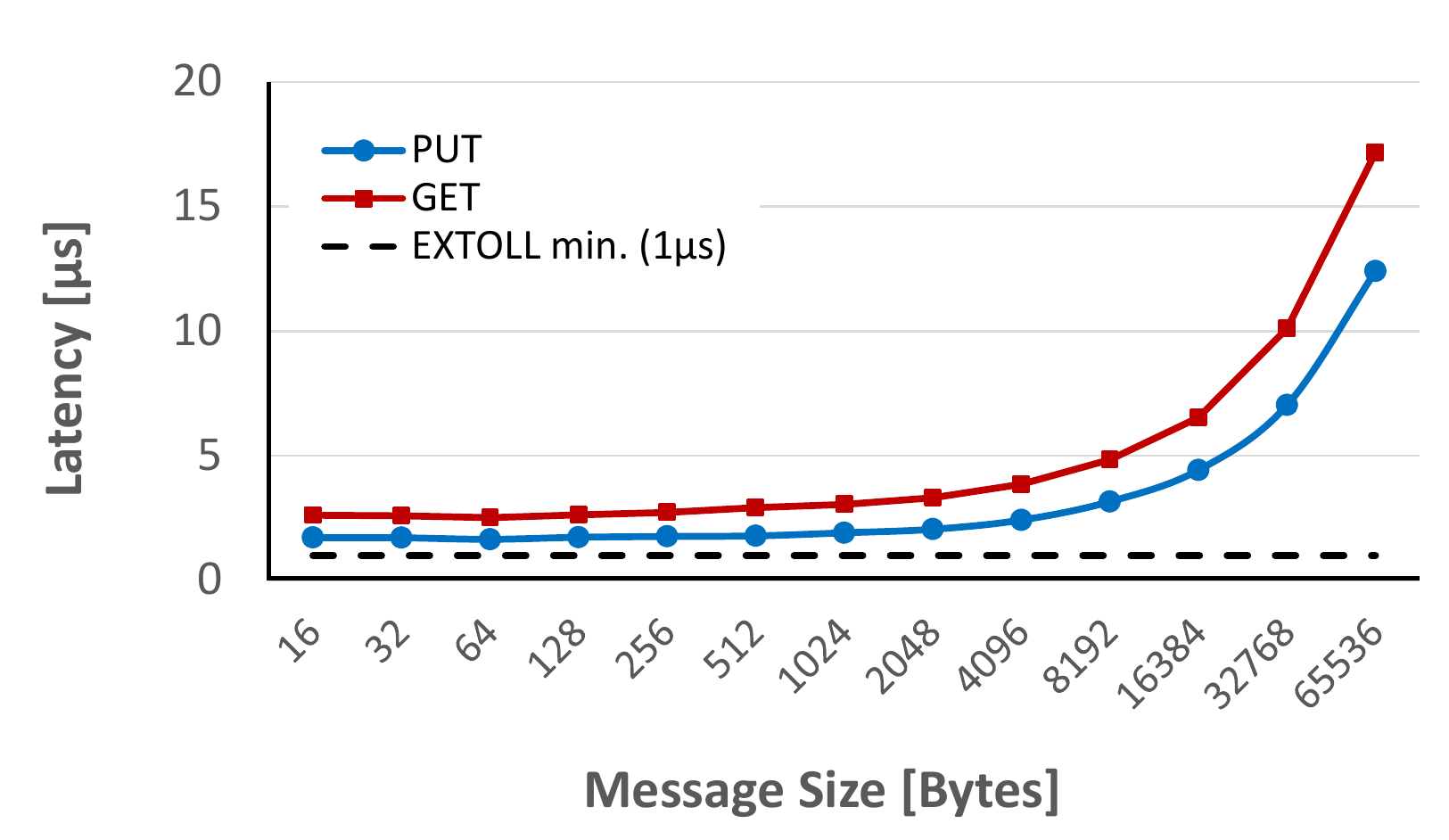}
\caption{RMA benchmarks (bandwidth and latency) on the NAM. 
	Best values achievable with Extoll are also plotted.}
\label{fig:nambench}
\end{figure}

The \textit{libNAM} library has been implemented to give system 
and application software access to the NAM memory 
pool~\cite{schmidtPhD}.
The library operates on top of the existing EXTOLL RMA API. 
The function calls provided by libNAM are very similar to EXTOLL's libRMA, 
such that existing applications that use the latter 
can be modified without much effort. 
Reading and writing is performed via send and receive buffers 
organized in a ring structure. The EXTOLL/NAM notification 
mechanism is used to handle the buffer space, i.e. 
to free up locations 
when data has been transmitted (\texttt{put}) 
or received (\texttt{get}). 
\figurename~\ref{fig:nambench} presents bandwidth and latency 
measurements applying these operations on the NAM, 
for various message sizes. 
Latency and bandwidth results to read and write data from and to the NAM 
are very close to the best achievable values on the network alone.


As a first NAM use-case, an improved  checkpointing/restart functionality 
has been implemented. It utilizes the NAM's FPGA to pull the required 
data from the compute nodes and to calculate parity 
information meant to be stored on the device locally. By this means
most of the checkpointing overhead is
offloaded to the NAM.
At the same time this is transparent to the applications enabled to
use SCR.
Results obtained from applications using this
functionality are presented in Section~\ref{sec:resil_results},  
\figurename~\ref{fig:xPic_NAM}.

\section{Software Environment}
\label{sec:sw}
The main guiding principle in the
software development for the Cluster-Booster concept has been
to stick, as much as possible, to standards and well 
established APIs.
The specific software features required to operate
Cluster and Booster as a unified system are implemented in
the lower layers of the software stack and are as transparent as
possible to application developers. In the end they experience a
software environment very similar to the one found
on any other current HPC system. By this means they do not have to
deal with the
underlying hardware complexity. Furthermore, their codes stay
portable and can run practically out-of-the-box on this new 
kind of platform, as well as on any other HPC system.

\subsection{Programming Environment}
\label{sec:progenv}
The ParaStation~MPI library has been specifically optimized
to efficiently run within both, Cluster and Booster, and also across
them. This global~MPI provides an efficient 
way of exchanging data between the two parts of 
the system~\cite{eicker:CCPE}.
It implements a heterogeneous,
\textit{global MPI} by exploiting semantic concepts long existing in
the MPI-standard. In particular, the MPI-2 function 
\texttt{MPI\_Comm\_spawn} realizes the offloading mechanism, 
which allows spawning groups of processes from Cluster to Booster 
(or vice-versa). 

\subsection{OmpSs abstraction layer}
\label{sec:ompss}
Directly employing the 
\texttt{MPI\_Comm\_spawn} function call forces the 
programmer to coordinate
and manage two or more sets of parallel MPI processes.
This includes the explicit handling of the required data exchanges
between Cluster
and Booster. This approach may become cumbersome 
for large and complex applications. To reduce the porting 
effort, an abstraction layer employing the global~MPI 
has been implemented already in the DEEP project. 
It is based on the OmpSs 
data-flow programming model~\cite{duran}. OmpSs enables 
application developers offloading large, complex tasks by 
simply annotating those parts of their 
code that shall run on a different part of the system via
\texttt{pragmas}~\cite{sainz}.

\subsection{I/O}
\label{sec:io}
The non-volatile memory of the \mbox{DEEP-ER} prototype 
(see Section~\ref{sec:nvm}) is used as the foundation of a scalable 
I/O infrastructure. The resulting software platform 
combines the parallel I/O library SIONlib~\cite{Frings:4447} 
with the  
parallel file system BeeGFS~\cite{beegfs}. Together, they enable
the efficient 
and transparent use of the underlying hardware and provide 
the functionality and performance required by data-intensive 
applications and multi-level checkpointing-restart techniques.

The I/O library SIONlib acts as a 
concentration-layer for applications employing task-local I/O. It aims
for the most efficient
use of the underlying parallel file system. For this, SIONlib bundles together
all data locally generated by applications, and stores 
it into one or very few large files, which the parallel file system
manage more easily. Furthermore, in \mbox{DEEP-ER} SIONlib bridges 
between the I/O and resiliency components of the 
software stack (Section~\ref{sec:chkp}). 
%

The file system utilized in \mbox{DEEP-ER} is 
BeeGFS. It provides a solid, common basis 
for high-performance, parallel I/O operations. Advanced 
functionality, such as a local cache layer in the file system, 
have been added to BeeGFS during the \mbox{DEEP-ER} project.
The cache domain --~based on BeeGFS on demand (BeeOND)~\cite{beeOnd}~-- 
stores data in fast node-local NVM devices 
and can be used in a synchronous or asynchronous mode. 
This speeds up the applications' I/O operations 
(Section~\ref{sec:io_results}) and reduces 
the frequency of accesses to the global storage, increasing 
the overall scalability of the file system.

\subsection{Resiliency}
\label{sec:resil}
The \mbox{DEEP-ER} project has adopted an improved 
application-based checkpoint-restart approach, 
in combination with a task-based resiliency strategy.

\subsubsection{Checkpoint/Restart}
\label{sec:chkp}
The Open Source Scalable Checkpoint-Restart 
library (SCR) offers a flexible interface 
for applications to write checkpoint information and to restart from 
those checkpoints in case of failure~\cite{moody:2010}. 
The user simply calls SCR 
and indicates the data required by the application 
to restart execution. This library keeps a database 
of checkpoints and their locations in preparation for 
eventual reinitializations. 

Combining SCR features with project-specific
software and hardware developments, four checkpoint/restart 
strategies can be employed in \mbox{DEEP-ER}, listed here ordered from
most basic to more advanced:

\begin{itemize}

\item \textbf{Single}: The \texttt{SCR\_SINGLE} feature stores 
checkpoints locally on each node. This enables applications
to recover from transient errors.

\item \textbf{Buddy}: This enhancement of the SCR's \textit{Partner} 
mode combines SCR, ParaStation~MPI, SIONlib, and BeeGFS. 
The standard \texttt{SCR\_PARTNER} mode first saves checkpoint-data
to the local node, re-reads it from there, sends it to a
\textit{Partner} node and writes it to its local storage.
This strategy enables
application to restart even after node-failures, recovering data 
from the companion node. Thus, the application can continue 
with minimum time-loss.
The \mbox{DEEP-ER} \textit{Buddy} enhancement utilizes
SIONlib to skip 
the intermediate step of reading the data from the local storage before 
sending it to the partner node, reducing the checkpointing overhead. 
While SCR in this approach keeps track of the association between host
nodes and buddies, SIONlib takes care that all 
MPI processes running on a single node jointly write their 
checkpoint-data into a single file on the buddy-node. Finally,
BeeOND stores the data itself on the cache file system 
on the local NVMe, eventually transferring it asynchronously to the 
permanent global storage. 

\item \textbf{Distributed XOR}: Both \texttt{SCR\_PARTNER} 
and \mbox{DEEP-ER}'s \textit{Buddy} mode save the whole 
checkpoint-data twice. 
This obviously doubles the required amount of storage per node and generates 
significant overhead due to the additional writing time. An alternative 
strategy is to generate and store parity information, instead of 
copying the full checkpoints. SCR can perform this by combining
checkpoint data from different nodes by an XOR operation. In a second step it
distributes the resulting parity data over all nodes.
If one node fails, the missing checkpoint-data can be reconstructed by 
combining the parity data and the checkpoint data from the remaining
nodes. This method saves the checkpoint data on the
node-local NVMe (as in the \textit{Single} mode) and distributes 
and remotely stores only the parity data, which is much smaller than
the full checkpoints.

\item \textbf{NAM XOR}: The network-attached memory (NAM) technology
developed in \mbox{DEEP-ER} is an ideal vehicle to accelerate the 
\textit{Distributed~XOR} strategy. SCR and \mbox{SIONlib} 
call libNAM to trigger data collection and parity computations on the
NAM. Furthermore, XOR-data is stored in this central location.   
This enables application developers 
to transparently checkpoint/restart to/from the 
NAM without modifying their codes. The high read/write performance
of the NAM (\figurename~\ref{fig:nambench}) and its accessibility at 
network-speed from all nodes in the system, largely reduces the 
checkpointing overhead when compared to the previous mode. At the same
time it provides a similar level of resiliency as the \textit{Buddy} mode.

\end{itemize}

\begin{figure}[!tbp]
  \centering
  \includegraphics[width=0.9\columnwidth]{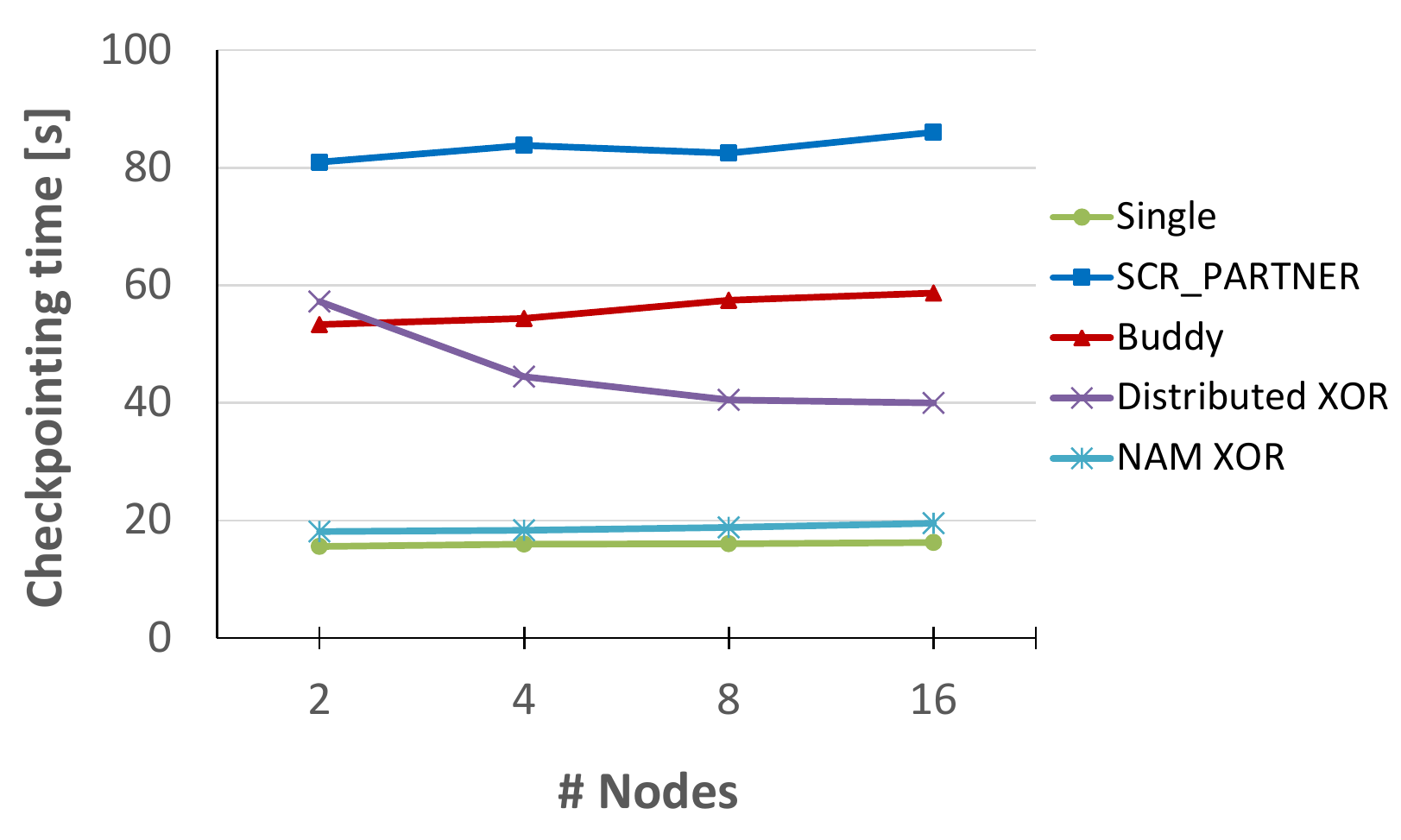}
    \caption{\label{fig:Nbody_chkp}
    N-body code testing various checkpointing strategies
    on the \mbox{DEEP-ER} Cluster (weak scaling).}
\end{figure}

Weak-scaling results obtained testing these functionalities 
on the \mbox{DEEP-ER} Cluster with the N-body code 
are presented in \figurename~\ref{fig:Nbody_chkp}. The \textit{Buddy}
and NAM~XOR methods developed in the \mbox{DEEP-ER} project are both faster than 
the equivalent SCR functions: \texttt{SCR\_PARTNER} and 
\textit{Distributed XOR}, respectively.
Results of a full application comparing the \textit{Distributed~XOR} 
with the \textit{NAM~XOR} modes are discussed in
Section~\ref{sec:resil_results}.

\subsubsection{OmpSs resiliency}
\label{sec:ompss_resil}
Also the OmpSs programming model has been enhanced by
three new resiliency features.

\begin{itemize} 
\item \textbf{Lightweight task-based checkpoint/restart mechanism}
writes the input of the OmpSs tasks into main memory
before starting them. Thus, they can be restarted in case of failure. 
If no error occurs and the task finishes successfully, the checkpoint is 
evicted.

\item \textbf{Persistent task-based checkpointing} saves all input
  dependencies
of a task. When the application is restarted after a crash, OmpSs 
transparently identifies the execution as a recovery and fast-forwards 
it to the point where the failure occurred, restoring the 
appropriate data. 

\item \textbf{OmpSs resilient offload}
is applied specifically to the offload mechanism developed 
in the DEEP projects (Section~\ref{sec:ompss}). 
The ParaStation process management daemon has been extended by an 
interface to query resiliency-related status information 
from the MPI layer and thus also from the OmpSs runtime 
environment. ParaStation~MPI itself is now able to detect, 
isolate and clean up failures of MPI-offloaded tasks, 
which can be then independently restarted without 
requiring a full application recovery. This enables
to recover failed offloaded tasks without losing the work 
that had been performed in parallel by other OmpSs tasks.
\end{itemize}

Application benchmarks utilizing the third OmpSs resiliency approach are 
described in Section~\ref{sec:resil_results}.

\section{Co-design applications}
\label{sec:apps}
Seven real-world HPC applications were 
chosen to steer and evaluate the design of the \mbox{DEEP-ER} 
hardware and software developments (Sections~\ref{sec:arch} 
and \ref{sec:sw}), and to benchmark their functionality 
and performance. The \mbox{DEEP-ER} applications come 
from a wide range of scientific areas, representing the typically 
broad user portfolio of a large-scale HPC center.
For the sake of brevity, details are given here only 
for three of the codes. Their 
results (Section~\ref{sec:results}) cover 
almost all the I/O and
resiliency features developed in \mbox{DEEP-ER}:

\begin{itemize}
\item \textbf{xPic} is a
  simulation code from KU~Leuven (Katholieke Universiteit Leuven) to 
  forecast space weather related events like e.g. damage of spacecraft 
  electronics or GPS signal scintillation. It simulates the
  inter-planetary plasma using the 
  Moment-Implicit method~\cite{ipic3d}. 
  Like most particle-in-cell 
  codes, xPic consists of two parts, a particle solver 
  and a field solver: The particle solver calculates the 
  motion of charged particles in response to the 
  electromagnetic field and the gathering of their 
  moments (e.g. net current and charge density); 
  the field solver computes the electromagnetic field 
  evolution in response to the moments. 
%
%

\item \textbf{GERShWIN}~\cite{app:gershwin} assesses
  human exposure to electromagnetic fields and is provided by Inria 
  (Institut National de Recherche en Informatique et en 
  Automatique). This application uses a Discontinuous 
  Galerkin - Time Domain solver 
  of the 3D Maxwell-Debye equation system~\cite{maxwell} 
  to simulate the propagation of electromagnetic waves 
  through human tissues. This field of research 
  studies for instance the effect of wireless communication 
  devices on head tissues or the implantation of antennas 
  in the human body for the purpose of monitoring 
  health-related devices. 
  

\item \textbf{FWI}~\cite{app:FWI} is a seismic imaging 
  code developed by the Barcelona Supercomputing Center (BSC). 
  Seismic imaging uses sound waves 
  to acquire the physical properties of the 
  subsoil from a set of seismic measurements. The 
  application starts from a guess (initial model) of the 
  variables (e.g., sound transmission velocity), 
  the stimulus introduced, and the recorded 
  signals. For the inversion several phases of iterative 
  computations (frequency cycles) are done until 
  the real value of the set of variables being inverted 
  is reached (with an acceptable error threshold). 
\end{itemize}

Further applications used in \mbox{DEEP-ER} are:
\begin{itemize}

\item \textbf{SKA data analysis pipeline} by ASTRON (Netherlands
  Institute for Radio Astronomy).

\item \textbf{TurboRvB} from CINECA 
  (Consorzio Interuniversitario del Nord-Est per il 
  Calcolo Automatico).

\item \textbf{SeisSol}
  from LRZ (Leibniz-Rechenzentrum der Bayerischen 
  Akademie der Wissenschaften).

\item \textbf{CHROMA}: 
  by the University of Regensburg.
\end{itemize}  

The role of the applications in \mbox{DEEP-ER} is two-fold: 
on the one hand, their requirements have provided co-design 
input to fix the characteristics of hardware and software 
components; on the other hand, the codes have evaluated the
project developments by running different uses cases on the 
\mbox{DEEP-ER} prototype. 
Examples of the co-design influence are the determination of the 
amount of memory to be available per node, the required MPI 
functionality when offloading code from one side to the 
other of the system, or the way in which the NAM should be 
addressed. A selection of the application results 
achieved by the first three codes (xPic, GERShWIN, and FWI) is
presented in the next section.

\section{Results}
\label{sec:results}
The hardware and software concepts developed in 
\mbox{DEEP-ER} have been evaluated using the co-design 
applications. Unless stated otherwise, all measurements 
have been obtained from the \mbox{DEEP-ER} prototype 
(Section~\ref{sec:sysconf}). The codes have been 
used in various simulation scenarios in order to test 
different system features (e.g. 
input parameters leading to more data communication were 
used to stress I/O features). 

\subsection{I/O application results}
\label{sec:io_results}
The features described in Section~\ref{sec:io} have lead to
several I/O improvements in the \mbox{DEEP-ER} 
applications. Some of them are shown here since the I/O
capabilities have a direct impact on checkpointing 
performance. The setup of all I/O experiments in this section 
is described in Table~\ref{tab:IOsetup}.

\begin{figure}[h!]
  \centering
  \includegraphics[width=0.9\columnwidth]{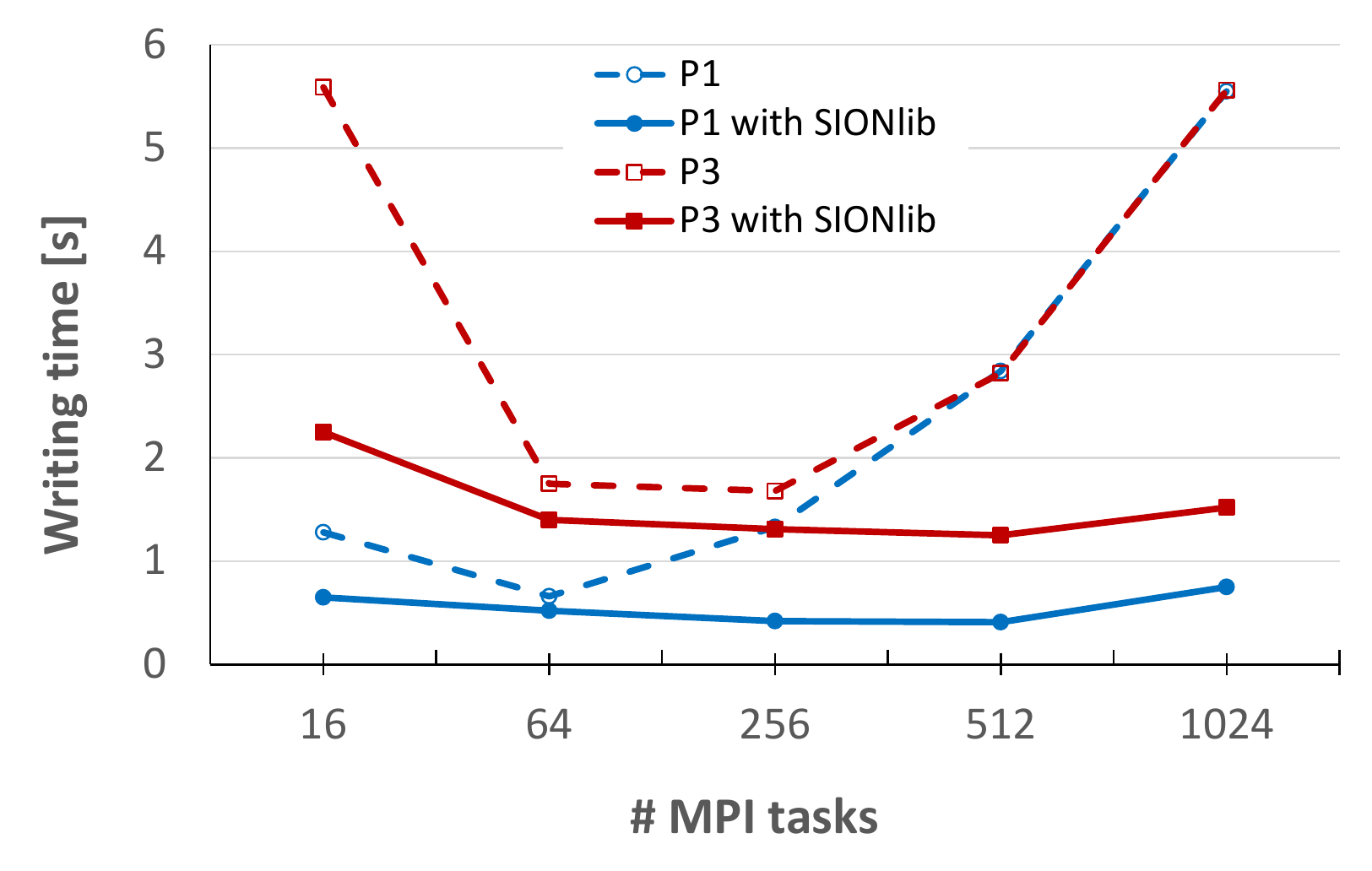}
  \caption{I/O improvement through SIONlib measured with GERShWIN.}
  \label{fig:xPic_SIONlib}
\end{figure}

\figurename~\ref{fig:xPic_SIONlib} shows the 
reduction of data writing time for the GERShWIN application when using SIONlib
to collectively carry out task-local I/O operations  
into a reduced number of files. 
Different use cases where tested, varying the Lagrange 
order of the calculations (order three (P3)
requires more data and provides
higher precision than order one (P1)). Significant 
performance improvements are achieved when using SIONlib: 
up to 7.4$\times$ faster for P1, and up to 3.7$\times$ 
for P3.

Even with the help of SIONlib, using the file system to 
execute I/O operations to the global storage from a large 
number of compute nodes may still lead
to a bottleneck at the storage: once the maximum 
storage bandwidth is reached, the bandwidth per node 
decreases when additional nodes participate in I/O.
In \mbox{DEEP-ER} this I/O scalability issue is targeted by
the BeeGFS caching-layer, which transparently employs the node-local
NVMe devices
(Section~\ref{sec:nvm}) as scalable local storage.
This has the effect of a constant storage bandwidth per node, 
significantly increasing the I/O performance and 
scalability of applications. 

\begin{figure}[h!]
  \centering
  \includegraphics[width=0.9\columnwidth]{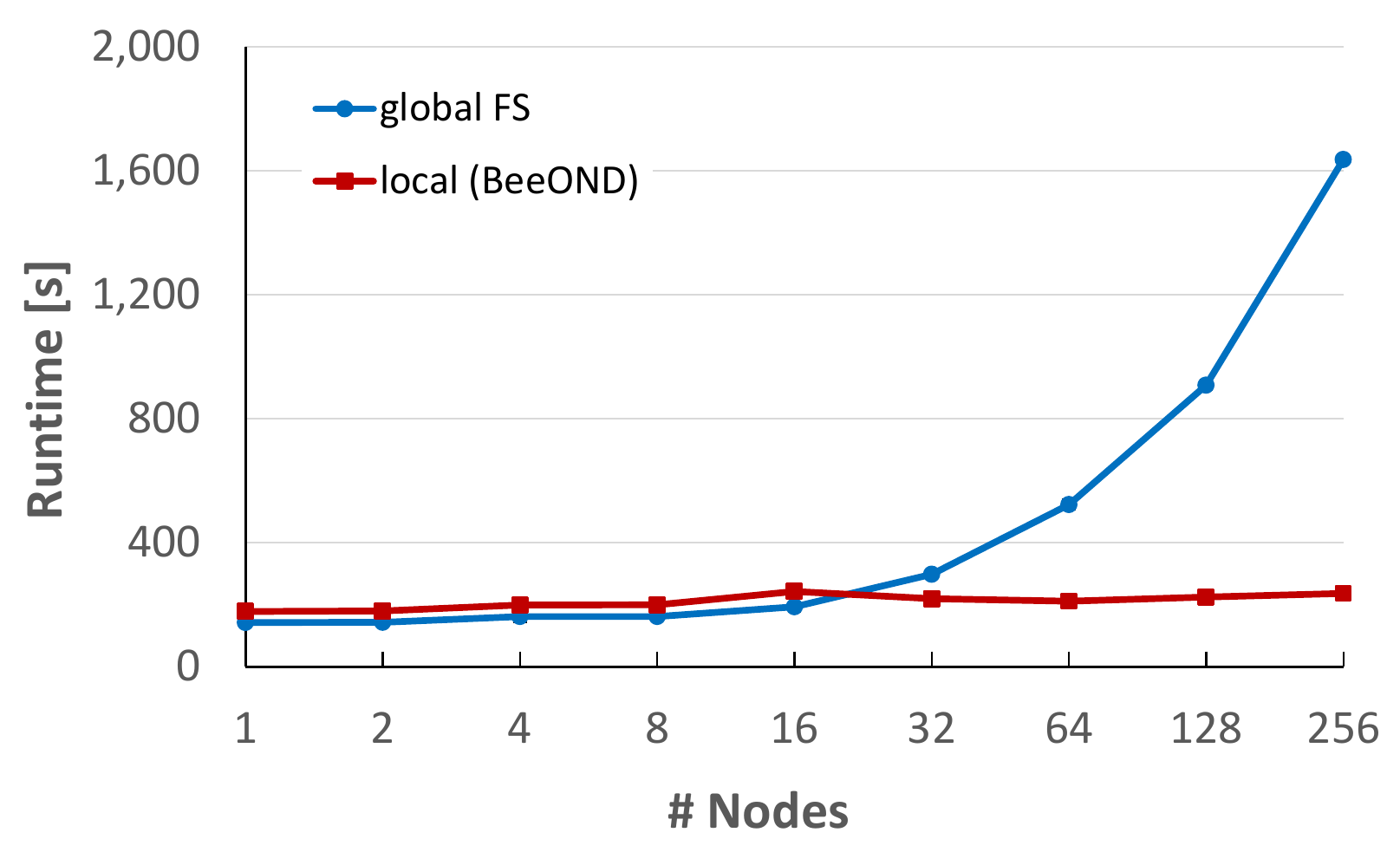}
  \caption{xPic on QPACE3: writing on global file system vs. 
  	using node-local storage with BeeOND.}
  \label{fig:xPic_NVM_QPACE3}
\end{figure}

Due to the small size of the \mbox{DEEP-ER} prototype, 
such scaling effects had to be measured on an alternative 
platform. The QPACE3~system
was selected, which is a Booster-like platform with 
672~KNL nodes~\cite{qpace3}, large enough to 
perform scalability measurements. 
\figurename~\ref{fig:xPic_NVM_QPACE3} shows
weak-scaling studies with xPic on QPACE3.

A software configuration similar
to the \mbox{DEEP-ER} platform could be installed on QPACE3.
Since this system lacks node-local NVMe devices, these had 
to be emulated by RAM-disks residing in the local memories of each node.
The absolute performance numbers are not comparable to NVMe 
(RAM on KNL is 75$\times$ faster than NVMe) but the 
advantage of using local storage devices with respect to the global 
storage system is clearly demonstrated, as well 
as the evident gain in performance when increasing 
the number of nodes. In fact, the application scales almost 
perfectly when using local storage.
This makes the application 7$\times$ faster compared to
writing directly to QPACE3's global file system which is also BeeGFS.

\begin{figure}[h!]
  \centering
  \includegraphics[width=0.9\columnwidth]{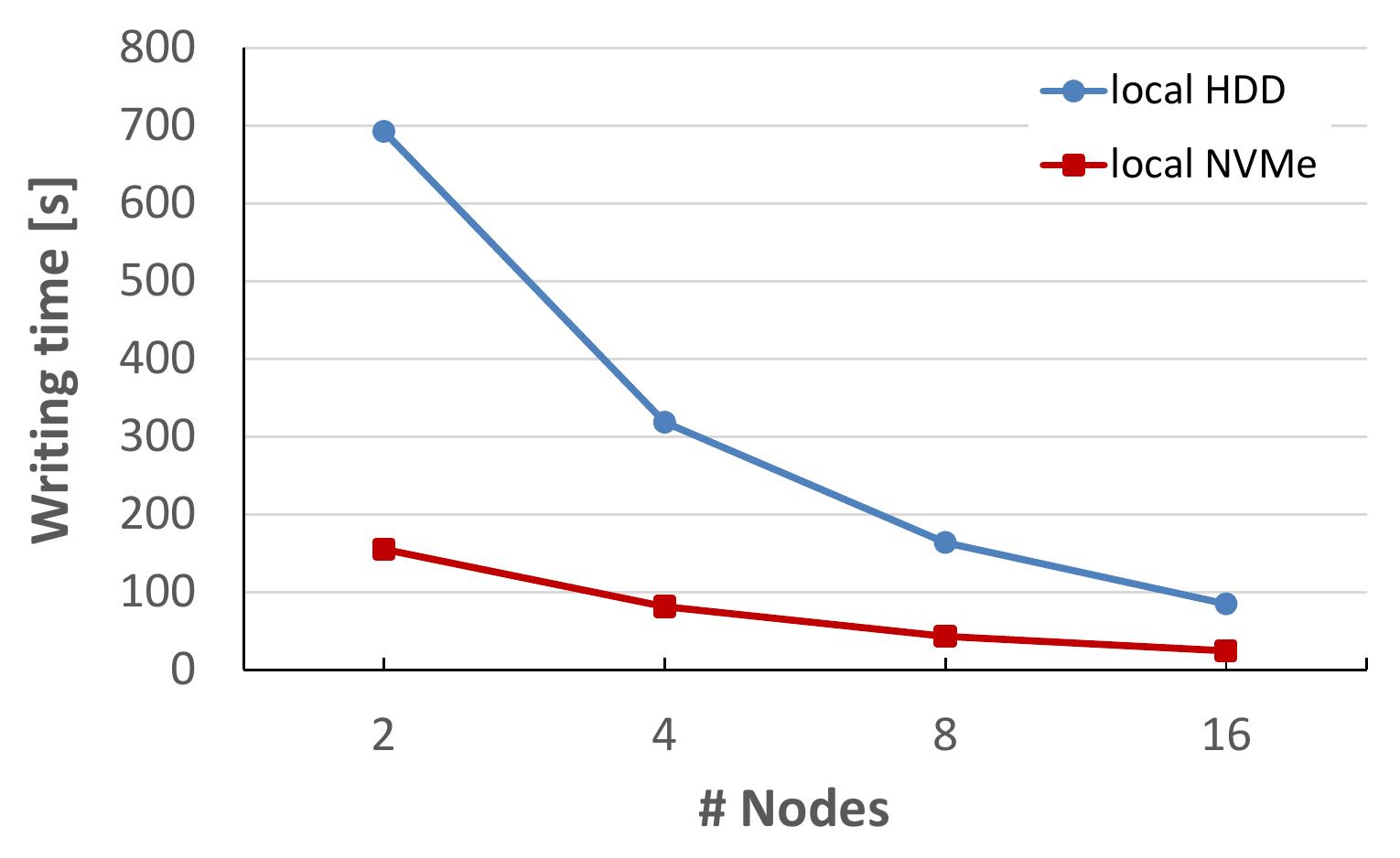}
  \caption{I/O operations to node-local NVMe vs. node-local
    HDD storage, measured by xPic on the \mbox{DEEP-ER} Cluster.}
  \label{fig:xPic_NVM_SDV}
\end{figure}

As demonstrated on QPACE3, the concept of writing to
node-local storage does not necessarily require NVMe devices 
to be attached to the compute nodes. The same strategy 
can be employed with other node-local storage technologies,
leading to different absolute performance numbers.
\figurename~\ref{fig:xPic_NVM_SDV} presents xPic measurements  
on the \mbox{DEEP-ER}~Cluster. Here both
NVMe and hard disks (HDD) are attached to each node.
Writing to the NVMe storage 
is up to 4.5$\times$ faster than writing to the 
node-local HDD. The absolute performance gain depends on the 
number of nodes performing I/O and, as explained above, 
the actual benefit from the NVMe-based local storage actually 
manifests when the node-count is very large.

\begin{table}[t]\scriptsize
  \centering
  \caption{\label{tab:IOsetup} Experiment setup used 
  in GERShWIN and xPic
  during the I/O measurements.}
  \begin{tabular}{|l|c|c|c|}
    \hline
    Experiment & \textbf{GERShWIN} & \textbf{xPic on QPACE3} & \textbf{xPic on DEEP-ER} \\
    \hline
    Data per CP & \parbox[c]{1.2cm}{3~GB (P1) \\ 6.6~GB (P3)} & 10 GB per node & 8 GB \\
    \hline 
    \# of CPs & 1 & 2 & 11 \\
    \hline
  \end{tabular}
\end{table}

\subsection{Resiliency}
\label{sec:resil_results}
During the \mbox{DEEP-ER} project multiple resiliency 
features have been developed (Section~\ref{sec:resil}). 
The setup of the resiliency experiments described here 
can be found in Table~\ref{tab:res_setup}.

\begin{figure}[h!]
  \centering
  \includegraphics[width=0.9\columnwidth]{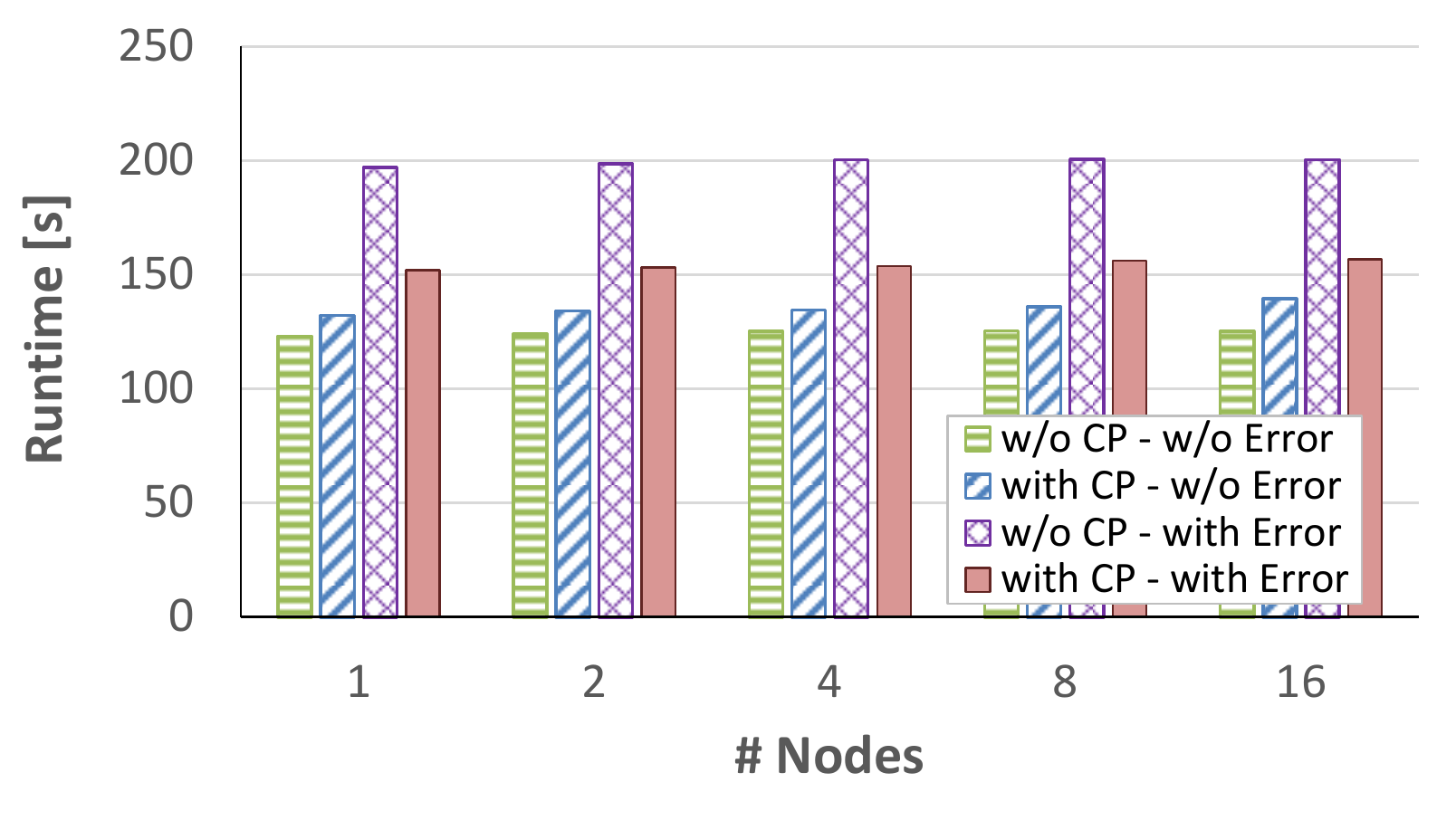}
  \caption{xPic testing \texttt{SCR\_PARTNER}. Tests done
  	writing  checkpoints (with CP) or not (w/o CP), for 
  	runs when an error occurs (with) or not (w/o).}
  \label{fig:xPic_SCR}
\end{figure}

\figurename~\ref{fig:xPic_SCR} displays the overhead 
and benefit of using the SCR library to save checkpoints (CP)
on the node-local NVMe (\texttt{SCR\_PARTNER}). An xPic benchmark
was selected that executes 100 iterations in the simulation. 
The benchmark was run with and without \texttt{SCR\_PARTNER}. 
In the latter case checkpoints are written every 10 iterations.  
Two error scenarios were tested: the first completes 
without error; in the second an error happens after 60 
iterations and then the application is restarted and runs through.
The measurements show that the overhead incurred by writing 
checkpoints with SCR is in average only 8\%, 
while it saves 23\% of the execution
time if a failure occurs according to the scenario.

\begin{figure}[!tbp]
  \centering
  \includegraphics[width=0.9\columnwidth]{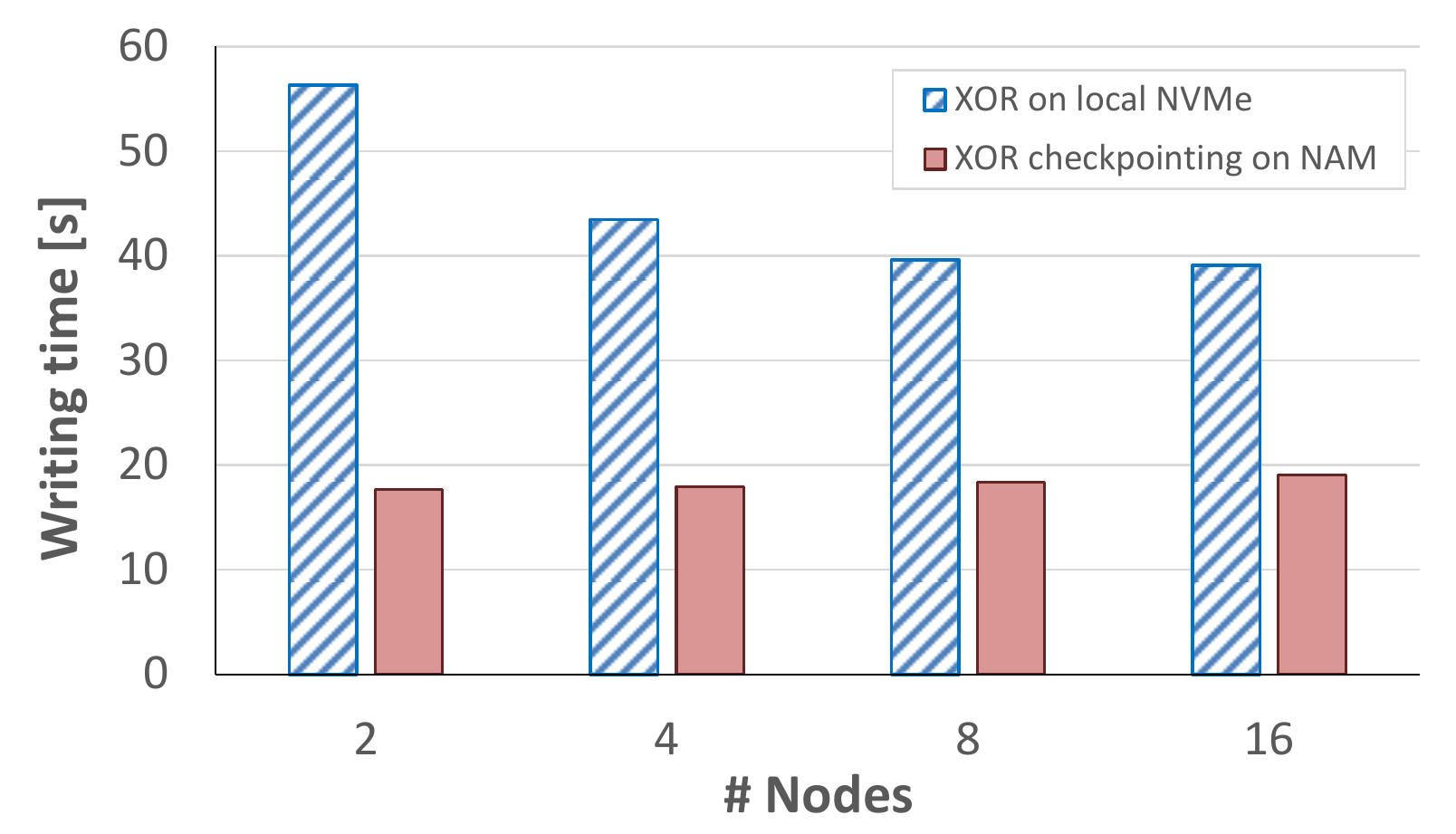} 
  \caption{\label{fig:xPic_NAM} \textit{Distributed XOR} vs. 
  \textit{NAM XOR} checkpointing strategies, 
  evaluated with xPic.}
\end{figure}

The checkpointing overhead can be further reduced 
calculating parity data (Section~\ref{sec:chkp}). 
\figurename~\ref{fig:xPic_NAM} 
shows a comparison between the \textit{Distributed XOR} and the
\textit{NAM XOR} checkpointing strategies. 
The latter realizes an up to 3$\times$ higher 
bandwidth, and leads to much better writing times: 
between 50\% and 65\% of time is saved when
storing XOR data to the NAM instead of storing it to the node-local
NVMe devices.


\begin{figure}[h!]
  \centering
  \includegraphics[width=0.9\columnwidth]{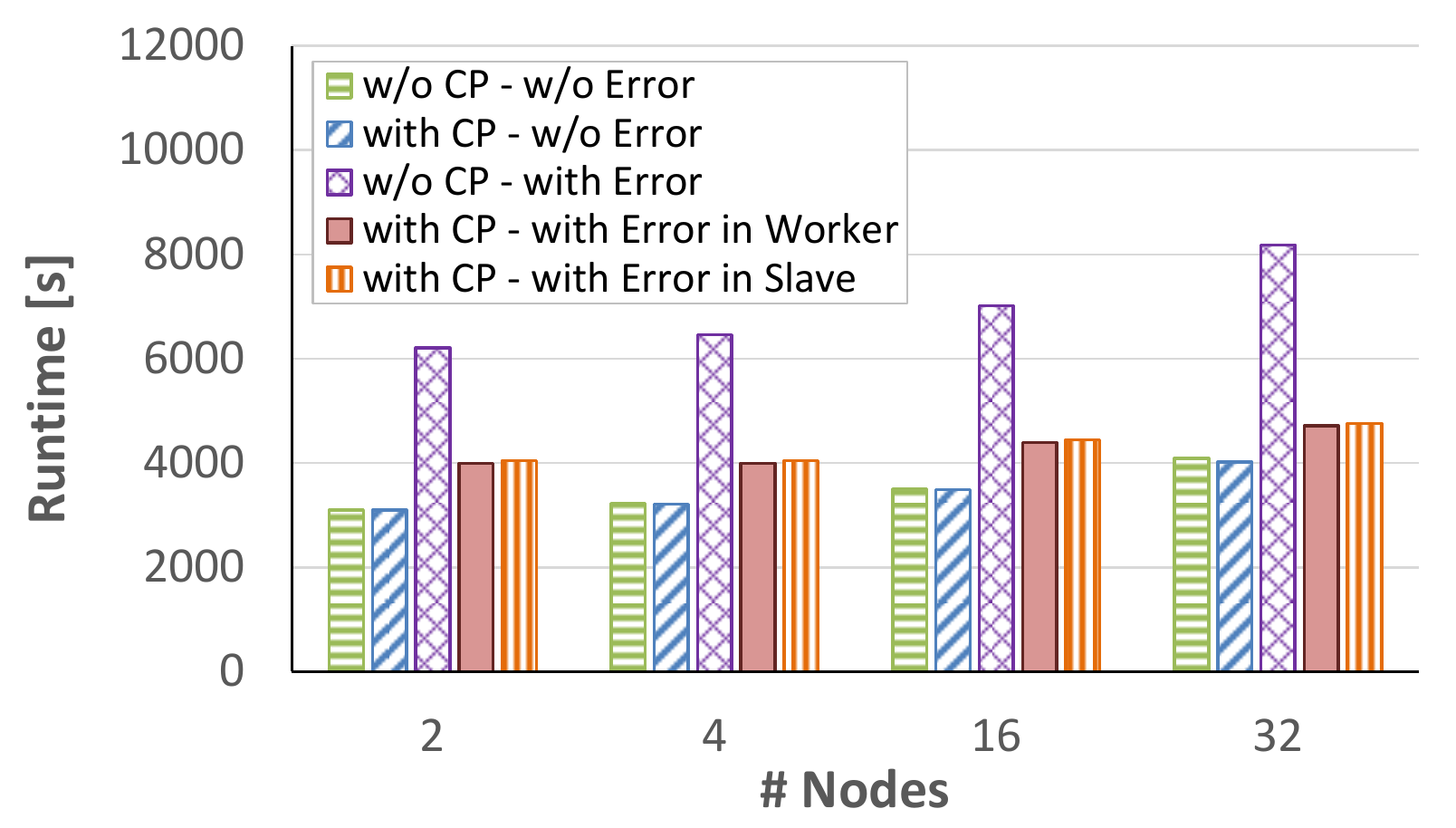}
  \caption{OmpSs task-based resiliency tested with FWI (with and without (w/o) 
  	checkpoints, with error -~in worker or slave~- and w/o errors).}
  \label{fig:FWI_OmpSs_resil}
\end{figure}

An alternative strategy applied in \mbox{DEEP-ER} to 
increase the applications' robustness against system 
failures is the OmpSs-offload resiliency functionality 
(Section~\ref{sec:ompss_resil}). 
\figurename~\ref{fig:FWI_OmpSs_resil} shows the results 
achieved when testing this approach with the FWI code,
on an Intel Sandy Bridge cluster (MareNostrum~3, at BSC). 
An error occurring right before the end of the execution 
nearly doubles the FWI runtime if no resiliency technique 
is activated. The new OmpSs feature enables up to 42\% 
time savings (an only 15\% longer execution when compared 
to a run without failures) and its overhead is 
negligible ($<$1\%).

\begin{table}[!htbp]\footnotesize
  \centering
  \caption{\label{tab:res_setup} Experiment setup for the resiliency measurements.}
  \begin{tabular}{|l|c|c|c|}
      \hline
      Experiment & \textbf{xPic SCR} & \textbf{xPic NAM} & \textbf{FWI} \\
      \hline
      Processed data & 32 GB per node & 20 GB per node  & 1 GB per node\\
      \hline 
      Data per CP & 8 GB & 2 GB & \\
      \hline 
      \# of CP & 4 & 10 &\\
      \hline 
    \end{tabular}
\end{table}

\section{Related work}
\label{sec:relwork}
The work on resiliency presented in this paper is based on the 
Scalable Checkpoint-Restart library (SCR)~\cite{moody:2010}. 
During the \mbox{DEEP-ER}
project a tight integration of SIONlib~\cite{Frings:4447} and SCR
was established. Task-local I/O as done by SCR typically 
uses many independent files. SIONlib concentrates
them into single or few shared files on a parallel file-system, 
what makes this type of I/O much more efficient. 
In \mbox{DEEP-ER} SIONlib is used as an abstraction-layer for both, buddy
checkpointing (similar to SCR\_PARTNER) and NAM
integration --~which might be seen as an hardware acceleration of
SCR's XOR checkpointing feature. Both approaches significantly
improve the I/O performance by preventing unnecessary read operations 
when creating partner and XOR checkpoints.

A similar approach as SCR is realized in the Fault Tolerance
Interface~(FTI)~\cite{fti}. In the meantime an effort was started to
create a common abstraction of SCR and FTI, to help
application developers avoiding code-adaptations to 
the specific checkpointing tool installed on a given
HPC system.

Transparent system level checkpointing is realized in Berkeley Lab Checkpoint
Restart (BLCR)~\cite{blcr}. In this approach applications do not have
to be modified in order to checkpoint them like it is necessary in
SCR or FTI, where explicit store and load operations have to be
introduced into the codes in order to create checkpoints with
all relevant data. The drawback of transparent checkpointing 
is that checkpoint wills grow much larger, since
the whole memory of the application has to
be dumped onto the underlying storage system. Furthermore, this type of
checkpointing is harder to implement since 
all MPI communication has to be brought into a globally consistent
state in order to get properly checkpointed, too.

The same approach as BLCR is used by the Distributed MultiThreaded
Checkpointing (DMTCP) efforts~\cite{dmtcp} therefore sharing the pros
and cons.

\section{Conclusions and Outlook}
\label{sec:concl}
The \mbox{DEEP-ER} project has introduced several
hardware and software
innovations to improve the I/O and resiliency capabilities 
of the Cluster-Booster architecture, and of HPC systems in general.
The central component is a multi-level memory hierarchy 
employing non-volatile memory (NVM) devices, locally attached to 
each of the nodes in the system.

The \mbox{DEEP-ER} I/O software system 
combines proven file-systems and libraries with extensions 
that allow optimal exploitation of the capabilities of the
memory and storage pool. 
The project's resiliency strategy relies on the 
combination of complementary functions to recover
from different kinds of errors with reduced overhead. 
Building upon the infrastructure provided by the Scalable 
Checkpoint/Restart library SCR, the \mbox{DEEP-ER} extensions 
reduce the checkpointing overhead keeping the same level of
resiliency.  An example is the \textit{Buddy} checkpointing
functionality that employs SIONlib to optimize \texttt{SCR\_PARTNER}.
Application resiliency is achieved with even better performance 
employing the network-attached memory (NAM) technology developed in
\mbox{DEEP-ER}. This special ``memory node'' is globally accessible
from all nodes and enables calculating and storing parity data 
of application checkpoints much faster than if done on the nodes
themselves. 
The achieved improvements in performance and resiliency 
have been demonstrated with real-world applications. 

The Cluster-Booster architecture -- which was first prototyped in
the DEEP projects series -- has gone into production in the meantime. 
A KNL-based, 5~PFlop/s Booster has recently been attached to the 
JURECA Cluster, which runs at JSC in Germany since 2015~\cite{jureca}. 
The JURECA Booster is already available to users as stand-alone system.
The Cluster-Booster communication protocol that allows applications 
running distributed over both parts of JURECA will be available 
to users as soon as its optimized version is ready for production. 

The \mbox{DEEP-ER} project is now completed and successfully
evaluated by external reviewers. Building on its results, 
the successor (\mbox{DEEP-EST}) project  
generalizes the Cluster-Booster concept to create the
\textit{Modular Supercomputing architecture}~\cite{Suarez:844072}. 
It combines any number 
of compute modules into a single computing platform. Each compute 
module is (as the Cluster and the Booster) a system of a 
potentially large size, tailored to 
the specific needs of a given kind of applications. 
To demonstrate its
capabilities, a three-module hardware prototype will be built,
covering the needs of both HPC and high performance data analytics
(HPDA) workloads.

\section*{Acknowledgements}
\label{sec:ack}
The authors thank all members of the \mbox{DEEP-ER} consortium 
for their strong commitment in the project, 
which led to several results described in
this paper. Special gratitude goes to J.~Schmidt (University of
Heidelberg) for the NAM results, A.~Galonska (JSC) for 
buddy-checkpointing benchmarks, and 
S.~Rodr\'{i}guez (BSC) OmpSs resiliency tests with FWI.

Part of the research presented here has received funding 
from the European Community's FP7/2007-2013 and H2020-FETHPC 
Programmes, under Grant Agreement n$^\circ$~287530 (DEEP),
610476 (\mbox{DEEP-ER}), and n$^\circ$~754304 
(\mbox{DEEP-EST}). The present publication reflects only the 
authors' views. The European Commission is not liable for 
any use that might be made of the information contained therein. 




%


\end{document}